\documentstyle[preprint,aps]{revtex}
\begin{document}

\draft
\preprint{RPI-N92-1994}
\title{ SENSITIVITY TO  PROPERTIES OF THE PHI-MESON \\
IN THE  NUCLEON STRUCTURE IN  THE CHIRAL SOLITON MODEL$^*$}
\author{
Nimai C. Mukhopadhyay$^{**}$ and L. Zhang$^{**}$\\
Physics Department, Rensselaer Polytechnic Institute\\
Troy, NY 12180-3590\\}
\date{ July, 1994}
\maketitle

\begin{abstract}
The influence of the $\phi$-meson on the nucleon properties in the
chiral soliton model is discussed.

\end{abstract}

\pacs{PACS numbers: 11.10.L, 14.20.D, 14.40}


\section{Introduction }
Properties of the $\phi$-meson and its photo- and electroproduction
are of fundamental interest to CEBAF and its possible  future extension. The
quark model assigns $\phi$ an  $s\bar{s}$ structure,
thus forbidding the radiative
decay $\phi\rightarrow\pi^0\gamma $. Experimentally it is also
found to be suppressed,
yielding  a branching fraction  of $1.3\times 10^{-3}$.
However, $\phi\rightarrow
\rho\pi$ and $\phi\rightarrow\pi^+ \pi^- \pi^0$ are not suppressed at
all. Thus, it is possible to incorporate the widths of these decays
into the  framework of the chiral soliton model\cite{ref1},
by making use of a
specific model for the compliance with the OZI rule. Such a model is,
for example, the $\omega -\phi$ mixing model\cite{ref2}. Consequence
of this  in the context of a chiral soliton model, which
builds on the $\pi\rho\omega a_1 (f_1)$ meson effective Lagrangian\cite{ref1},
is the content of this report.
\section{ Results}
We summarize, in Table 1, nucleon observables computed in our
chiral soliton model, with and without the $\phi$-decay constraints
put in, by using the picture\cite{ref2}
\begin{equation}
\phi_{physical}=\phi +\varepsilon\omega
\end{equation}
where \begin{eqnarray}
\phi =s\bar{s}, & \omega =\frac{1}{\sqrt{2}}(u\bar{u}+d\bar{d}),
\end{eqnarray}
and $\varepsilon$ is  determined from the flavor SU(3) mass
splitting,
\begin{equation}
\varepsilon =0.060\pm 0.014.
\end{equation}

We notice that {\em the phenomenological outcome, for the soliton
properties is worse, for all soliton properties under
consideration, except mass of the soliton, when we take into
account the $\phi$- decay properties},  in Model B. The
nucleon properties computed here
are the leading order ones in the $(N_c)^{-1}$ expansion, $N_c$
being the number of colors in QCD.

\section{ Summary and Conclusions }
Our attempt to include the $\phi$- decay properties in the chiral soliton
model makes nucleon properties almost universally worse, if we take the
picture of $\phi$, discussed above, for the protection of the OZI rule.
This is not a failure of the chiral
soliton model, but an indication that {\em the OZI rule, in general,
and its influence on the $\phi$ decays, in particular,
are not understood at
present}, despite many brave attempts\cite{ref3}. The Wess-Zumino-Witten
sector of the chiral soliton model is sensitive to the way we preserve
this rule.

Thus, in the context of nucleon structure, the $\phi$-physics
here offers a reminder that we are far from solving the problem of
strange quark contents in ordinary (supposedly non-strange)
hadrons. We need to understand the $phi$ properties a great deal
better. CEBAF I and II could help in this enterprise.

\vglue 0.3cm
 \noindent
$^*$  {\small Presented by N. C. Mukhopadhyay}\\
$^{**}$  {\small Supported by the U. S. Department of
Energy}


\newpage

\begin{table}
\caption{\tenrm Nucleon observables calculated in our chiral soliton
model. Model A has no $\phi$- decay
constraint, B does.  Notations are standard and units are obvious.}
\label{table1}
\bigskip
\begin{center}
\begin{tabular}{ccccccccc}
&$M_H$ &$g_A$& $g_{\pi NN}$&$\mu_V$&$<r>^{I=0}_E$&$<r>^{I=1}_M$&
$r_A$ &$\sigma$\\
\hline
A&1379&0.90&13.30&2.06&0.91& 0.92& 0.62& 43.5\\
B&1070&0.59&6.74&1.22&0.72&0.76&0.45&23.0\\
Exp.&939& 1.26& 13.45& 2.35& 0.79& 0.87& 0.63& 45\\
&$\pm 1$&$\pm 0.01$&$\pm 0.05$&$\pm 0.00$&$\pm 0.01$&$\pm 0.07$&
$0.03$&$\pm 10$ \\
\end{tabular}
\end{center}
\end{table}


\newpage

\begin{thebibliography}{99}
\bibitem{ref1} L. Zhang, and N. C. Mukhopadhyay {\it Mod. Phys. Lett.}
{\bf A9}  (1994) 935, and {\it Phys. Rev.}{ \bf D}, in press.

\bibitem{ref2} P. Jain {\it et al.}, {\it Phys. Rev.} {\bf D37}  (1988)
3253; U.-G. Meissner {\it et al.}, {\it ibid.}{ \bf D39}  (1989) 1956;
{\bf D40}  (1989) 362(E).

\bibitem{ref3} See, for example, P. Geiger and
N. Isgur, {\it Phys. Rev. } {\bf D47}  (1993) 5050.

\end{thebibliography}
\end{document}